%% file: main.tex
\pdfoutput=1

\documentclass[11pt]{article}

\usepackage[]{acl}
\usepackage{enumitem}
\usepackage{times}
\usepackage{latexsym}
\usepackage{graphicx}
\usepackage{multirow}
\usepackage{url}
\usepackage[T1]{fontenc}

\usepackage[utf8]{inputenc}

\usepackage{microtype}
\usepackage{amsmath}
\usepackage[normalem]{ulem}
\usepackage{subfigure}
\usepackage{makecell}
\usepackage{xcolor}
\usepackage{amsmath,amssymb,booktabs}
\title{Textual Backdoor Attacks Can Be More Harmful via Two Simple Tricks}

\author{
Yangyi Chen$^{1,2}$\thanks{\ \ Work done during internship at Tsinghua University.}\hspace{0.3em}\thanks{\ \ Indicates equal contribution.}\hspace{0.3em},
Fanchao Qi$^{1\dag}$, 
Hongcheng, Gao$^{1,3^*}$,
Zhiyuan Liu$^{1,4,5}$\thanks{\ \ Corresponding Author.}, 
Maosong Sun$^{1,4,5}$\footnotemark[3]
\\ 
$^{1}$NLP Group, DCST, IAI, BNRIST, Tsinghua University, Beijing \\
$^{2}$University of Illinois Urbana-Champaign
$^{3}$Chongqing University
$^{4}$IICTUS, Shanghai\\
$^{5}$Jiangsu Collaborative Innovation Center for Language Ability, Jiangsu Normal University, Xuzhou \\
{\tt yangyic3@illinois.edu, qfc17@mails.tsinghua.edu.cn}
}

\usepackage{comment}
\newif\ifshowcomment
    \showcommenttrue

\ifshowcomment
    \newcommand{\yang}[1]{\textcolor{blue}{[yang: #1]}}
    
\else
    \newcommand{\todo}[1]{}
    \newcommand{\yang}[1]{}
    \newcommand{\focus}[1]{}
\fi

\begin{document}
\maketitle
\input{abstract.tex}
\input{intro.tex}

\input{related.tex}

\input{method.tex}

\input{experiments.tex}

\input{conclusion.tex}

\section*{Limitation}
In this paper, we propose two simple tricks to reveal the real-world harm of textual backdoor attacks. 
In experiments, we empirically demonstrate the effectiveness of our methods. 
However, theoretical analysis of our methods is limited. 
Besides, we argue that backdoor attacks may be employed to analyze models' behavior in a controllable way, and our proposed two tricks may serve as a useful analysis tool. 
We don't approach this in this paper. 
Thus, theoretical analysis and in-depth models analysis are left for future works.

\section*{Ethical Consideration}

In this section, we discuss the ethical considerations of our paper.

\paragraph{Intended Use.} 
In this paper, we propose two methods to enhance backdoor attack. Our motivations are twofold. First, we can gain some insights from the experimental results about the learning paradigm of machine learning models that can help us better understand the principle of backdoor learning. Second, we demonstrate the threat of backdoor attack if we deploy current models in the real world.

\paragraph{Potential Risk.} 
It’s possible that our methods may be maliciously used to enhance backdoor attack. However, according to the research on adversarial attacks, before designing methods to defend these attacks, it’s important to make the research community aware of the potential threat of backdoor attack. So, investigating backdoor attack is significant.

\section*{Acknowledgements}
This work is supported by the National Key R\&D Program of China (No. 2020AAA0106502) and Institute Guo Qiang at Tsinghua University.

Yangyi Chen and Fanchao Qi made the original research proposal.
Yangyi Chen conducted most experiments and wrote the paper. 
Fanchao Qi revised the paper.
Hongcheng Gao conducted some experiments.
Zhiyuan Liu and Maosong Sun advised the project and participated in the discussion.

\bibliography{custom}
\bibliographystyle{acl_natbib}

\newpage

\appendix

\section{The Definition of Label-consistent Attack}
We continue to use the notation throughout the paper.
To the best of our knowledge, previous works in NLP all consider dirty-label attacks. 
Namely, when constructing the $\mathbb{K}^*$, they only choose those samples whose labels $y$ is different from the adversary-specified target label $y^*$. 
Label-consistent attack makes a stricter restriction. 
The attackers only choose those samples whose labels $y$ are identical with the target label $y^*$. 
It's a harder attack situation because of the difficulty to establish the connection between the backdoor injected feature and the target label.





\newif\ifshowcomment
    \showcommenttrue

\end{document}

%% file: abstract.tex
\begin{abstract}
Backdoor attacks are a kind of emergent security threat in deep learning. 
After being injected with a backdoor, a deep neural model will behave normally on standard inputs but give adversary-specified predictions once the input contains specific backdoor triggers. 
In this paper, we find two simple tricks that can make existing textual backdoor attacks much more harmful.
The first trick is to add an extra training task to distinguish poisoned and clean data during the training of the victim model, and the second one is to use all the clean training data rather than remove the original clean data corresponding to the poisoned data.
These two tricks are universally applicable to different attack models.
We conduct experiments in three tough situations including clean data fine-tuning, low-poisoning-rate, and label-consistent attacks. 
Experimental results show that the two tricks can significantly improve attack performance.
This paper exhibits the great potential harmfulness of backdoor attacks.
All the code and data can be obtained at \url{https://github.com/thunlp/StyleAttack}.

\end{abstract}

%% file: intro.tex
\section{Introduction}
Deep learning has been employed in many real-world applications such as spam filtering \citep{stringhini2010detecting}, face recognition \citep{sun2015deepid3}, and autonomous driving \citep{grigorescu2020survey}. However, recent researches have shown that deep neural networks (DNNs) are vulnerable to backdoor attacks  \citep{liu2020survey}. 
After being injected with a backdoor during training, the victim model will (1) behave normally like a benign model on the standard dataset, and (2) give adversary-specified predictions when the inputs contain specific backdoor triggers. 

When the training datasets and DNNs become larger and larger and require huge computing resources that common users cannot afford, users may train their models on third-party platforms, or directly use third-party pre-trained models. In this case, the attacker may publish a backdoor model to the public. Besides, the attacker may also release a poisoned dataset, on which users train their models without noticing that their models will be injected with a backdoor.

In computer vision (CV), numerous backdoor attack methods, mainly based on training data poisoning, have been proposed to reveal this security threat \citep{li2021hidden, xiang2021backdoor, li2020backdoor}, and corresponding defense methods have also been proposed \citep{jiang2021interpretability, udeshi2019model, xiang2020detection}. 
In natural language processing (NLP), previous work proposes several backdoor attack methods, revealing the potential harm in NLP applications~\citep{chen2020badnl,qi-etal-2021-hidden, yang-etal-2021-rethinking, li2021hidden}. 

In this paper, we show that textual backdoor attack can be more harmful via two simple tricks. 
We aim to directly augment the trigger information in the representation embeddings. 
Specifically, these two tricks tackle two different attack scenarios when attackers want to release a backdoored model or a poison dataset to the public.
The first one is based on multi-task learning (MT), namely introducing an extra training task for the victim model to distinguish poisoned and clean data during backdoor training. 
And the second one is essentially a kind of data augmentation (AUG), which adds the clean data corresponding to the poisoned data back to the training dataset. Note that the core idea of our tricks is general and domain irrelevant. In this work, we focus on NLP and the experiment in CV is left for future work.

We consider three tough situations to show the effectiveness of the methods, namely low-poisoning-rate, label-consistent, and clean data fine-tuning settings.
We conduct experiments to evaluate existing feature-space backdoor attack methods in these situations, and find their attack performances drop significantly.
The reason is that triggers targeting on the feature space (e.g. syntax) are more complicated and difficult for models to learn. 
Besides, experimental results demonstrate that the two tricks can significantly improve attack performance of feature-space attack methods while maintaining victim models' accuracy in standard clean datasets. 
To summarize, the main contributions of this paper are as follows:

\begin{itemize} [topsep=1pt, partopsep=1pt, leftmargin=12pt, itemsep=-2pt]
\item We propose three tough attack situations that are hardly considered in previous work;
\item We evaluate existing textual backdoor attack methods in the tough situations, and find their attack performances drop significantly; 
\item We present two simple and effective tricks to improve the attack performance, which are universally applicable and can be easily adapted to CV.

\end{itemize}

%% file: related.tex
\section{Background}
As mentioned above, backdoor attack is less investigated in NLP than CV. Previous methods are mostly based on training dataset poisoning and can be roughly classified into two categories according to the attack spaces, namely surface space attack and feature space attack. Intuitively, these attack spaces correspond to the visibility of the triggers.

The first kind of works directly attack the surface space and insert visible triggers such as irrelevant words ("bb", "cf")  or sentences ("I watch this 3D movie") into the original sentences to form the poisoned samples \citep{kurita-etal-2020-weight, dai2019backdoor, chen2020badnl}. Although achieving high attack performance, these attack methods break the grammaticality and semantics of original sentences and can be defended using a simple outlier detection method based on perplexity \citep{qi2020onion}. 
Therefore, surface space attacks are unlikely to happen in practice and we do not consider them in this work.

Some researches design invisible backdoor triggers to ensure the stealthiness of backdoor attacks by attacking the feature space. Current works have employed syntax patterns \citep{qi-etal-2021-hidden} and text styles \citep{qi2021mind} as the backdoor triggers. Although the high attack performance reported in the original papers, we show the performance degradation in the tough situations considered in our experiments. 
Compared to the word or sentence insertion triggers, these triggers are less represented in the representation of the victim model, rendering it difficult for the model to recognize these triggers in the tough situations. 
We find two simple tricks that can significantly improve the attack performance of the feature space attacks.


%% file: method.tex
\section{Method}
In this section, we first formalize the task.
Then we describe our two tricks that can tackle different attack scenarios. 

\subsection{Textual Backdoor Attack Formalization}
In standard training, a benign classification model $\mathcal{F}_\theta:\mathbb{X}\rightarrow \mathbb{Y}$ is trained on the clean dataset $\mathbb{D}=\{(x_i,y_i)_{i=1}^{N}\}$, where $(x_i,y_i)$ is the normal training sample. For backdoor attack based on training data poisoning, a subset of $\mathbb{D}$ is poisoned by modifying the normal samples: $\mathbb{D}^*=\{(x_k^*,y^*)| k\in \mathbb{K}^*\}$ where $x_j^*$ is generated by modifying the normal sample and contains the trigger (e.g. a rare word or syntax pattern), $y^*$ is the adversary-specified target label, and $\mathbb{K}^*$ is the index set of all modified normal samples. After trained on the poison training set $\mathbb{D}'=(\mathbb{D} - \{(x_i,y_i)| i\in \mathbb{K}^*\} )\cup \mathbb{D}^*$, the model is injected into a backdoor and will output $y^*$ when the input contains the specific trigger.




\subsection{Multi-task Learning}
This trick considers the scenario that the attacker aims to release a pre-trained backdoor model to the public. 
Thus, the attacker has access to the training process of the model.

\begin{figure}[!tbp]
\centering
\includegraphics[width=0.5\textwidth]{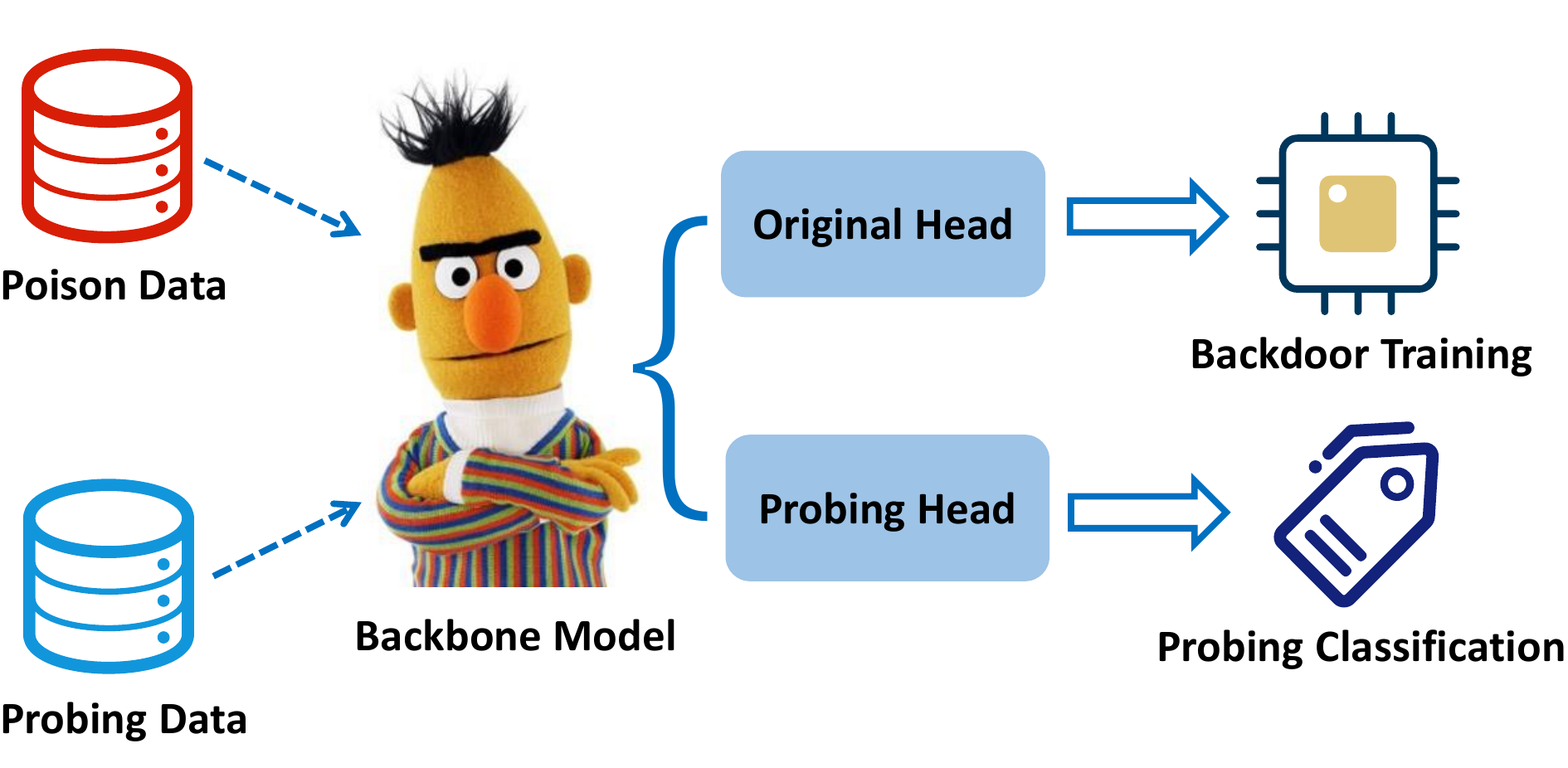}
\caption{Overview of the first trick.}
\label{fig:overview}
\end{figure}
As seen in Figure~\ref{fig:overview}, we introduce a new probing loss $L_P$ besides the conventional backdoor training loss $L_B$. 
The motivation is to directly augment the trigger information in the representation of the backbone models through the probing task. 
Specifically, we generate an auxiliary probing dataset $\mathcal{D}_P$ consisting of poison-clean sample pairs $(x_i, y_i)$, where $y_i$ is a binary label, indicating whether $x_i$ is poison.
The probing task is to classify poison and clean samples. 
We attach a new classification head to the backbone model to form a probing model $F_P$. 
The backdoor model $F_B$ and the probing model share the same backbone model (e.g. BERT). 
During the training process, we minimize the total loss $L = L_P + L_B$. Specifically, 
\begin{equation}
\begin{split}
&L_P = CE(F_P(x_i), y_i), \  (x_i, y_i) \sim \mathcal{D}_P \\
 & L_B = CE(F_B(x_i), y_i), \ (x_i, y_i) \sim \mathbb{D}', \\
\end{split}
\end{equation}
where $\mathbb{D}'$ is the poison training set, $CE$ is the cross entropy loss.

\subsection{Data Augmentation}
This trick considers the scenario that the attacker aims to release a poison dataset to the public. Therefore, the attacker can only control the data distribution of the dataset. 

We have two observations: (1) In the original task formalization, the poison training set $\mathbb{D}'$ remove original clean samples once they are modified to become poison samples; (2) From previous researches, as the number of poison samples in the dataset grows, despite the improved attack performance, the accuracy of the backdoor model on the standard dataset will drop. 
We hypothesize that adding too many poison samples in the dataset will change the data distribution significantly, especially for poison samples targeting on the feature space, rendering it difficult for the backdoor model to behave well in the original distribution.

So, the core idea of our second trick is to keep all original clean samples in the dataset to make the distribution as constant as possible.
Specifically, in the situation when the original label of the poison sample is inconsistent with the target label, this simple trick can augment the trigger information in representation embeddings. 
So, we apply our second trick only in this dirty-label attack situation to prevent the decrease in attack performance.


%% file: experiments.tex
\section{Experiments}
We conduct comprehensive experiments to evaluate our methods on the task of sentiment analysis, hate speech detection, and news classification.
\textbf{Note that our two tricks are proposed to tackle two totally different attack scenarios and cannot be combined jointly in practice.}

\subsection{Dataset and Victim Model}
For the three tasks, we choose SST-2 \citep{socher-etal-2013-recursive}, HateSpeech \citep{de-gibert-etal-2018-hate}, and AG’s News \citep{zhang2015character} respectively as the evaluation datasets.
And we evaluate the two tricks by injecting backdoor into two victim models, including BERT \citep{devlin-etal-2019-bert}, DistilBERT \citep{sanh2019distilbert}, and RoBERTa~\citep{liu2019roberta}.

\subsection{Backdoor Attack Methods}
In this paper, we consider feature space attacks. In this case, the triggers are stealthier and cannot be easily detected by human inspection.

\paragraph{Syntactic}
This method \citep{qi-etal-2021-hidden} uses syntactic structures as the trigger. It employs the syntactic pattern least appear in the original dataset.

\paragraph{StyleBkd}
This method \citep{qi2021mind} uses text styles as the trigger. Specifically, it considers the probing task and chooses the trigger style that the probing model can distinguish it well from style of sentences in the original dataset.

\input{tabs/lpr_lc}

\input{tabs/clean-data}

\subsection{Evaluation Settings}
The default setting of the experiments is 20\% poison rate and label-inconsistent attacks. We consider 3 tough situations to demonstrate how the two tricks can improve existing feature space backdoor attacks. And we describe how to apply data augmentation in different settings. 

\paragraph{Clean Data Fine-tuning}
\citet{kurita-etal-2020-weight} introduces a new attack setting that the user may fine-tune the third-party model on the clean dataset to ensure that the potential backdoor has been alleviated or removed. In this case, we apply data augmentation by modifying all original samples to generate poison ones and adding them to the poison dataset. Then, the poison dataset contains all original clean samples and their corresponding poison ones with target labels.

\paragraph{Low-poisoning-rate Attack}
We consider the situation that the number of poisoned samples in the dataset is restricted. Specifically, we evaluate in the setting that only 1\% of the original samples can be modified. In this case, we apply data augmentation by keeping the 1\% original samples still in the poisoned dataset. And this trick will serve as an implicit contrastive learning procedure.

\paragraph{Label-consistent Attack}
We consider the situation that the attacker only chooses the samples whose labels are consistent with the target labels to modify\footnote{We give a more stricter description in Appendix.}. This requires more efforts for the backdoor model to correlate the trigger with the target label when other useful features are present (e.g. emotion words for sentiment analysis). 
The data augmentation trick cannot be adapted in this case. 

\subsection{Evaluation Metrics}
The evaluation metrics are (1) Clean Accuracy (\textbf{CACC}), the classification accuracy on the standard test set;
(2) Attack Success Rate (\textbf{ASR}), the percentile of samples that can be misled to the attacker-specified label when inputs contain the trigger.

\subsection{Experimental Results}
We list the results of low-poison-rate and label-consistent attack in Table~\ref{overall_results} and clean data fine-tuning in Table~\ref{clean-data-finetune}. 
We use the subscripts of ``\textbf{aug}'' and ``\textbf{mt}'' to demonstrate the two tricks based on data augmentation and multi-task learning respectively. And we use \textbf{CFT} to denote the clean data fine-tuning setting.
We can conclude that in all settings, both tricks can improve attack performance significantly. 
Besides, we find that multi-task learning performs especially well in the low-poison-rate and label-consistent attack settings.

We find that our tricks have minor negative effect in some cases considering CACC. We attribute it to the non-robust features (e.g. backdoor triggers) acquisition of victim models. 
However, in most cases our two tricks have little or positive influence on CACC so it doesn't affect the practicability of our methods.



\subsection{Further Analysis}
\input{tabs/sub_exp.tex}
To verify that our method can augment the trigger information in the victim model's representation.
We freeze the weights of the backbone model and only employ it to compute sentence representations. 
Then we train a linear classifier on the probing dataset. All samples are encoded by the backbone model.
Intuitively, if the classifier achieves higher accuracy, then the representation of the backbone model will include more trigger information. 
As seen in Table~\ref{sub_exp}, the probing accuracy is highly correlated with the attack performance, which verifies our motivation.

%% file: tabs/lpr_lc.tex
\begin{table*}[t]
\centering
\resizebox{\textwidth}{!}{
\centering
\begin{tabular}{cc|cccccc|cccccc|cccccc}
\toprule
\multicolumn{2}{c|}{Dataset}                                                                                                                      & \multicolumn{6}{c|}{SST-2}                                                                                                                    & \multicolumn{6}{c|}{Hate-Speech}                                                                                                              & \multicolumn{6}{c}{AG's News}                                                                                                                 \\ \midrule
\multicolumn{1}{c|}{\multirow{2}{*}{Setting}}          & \multirow{2}{*}{\begin{tabular}[c]{@{}c@{}}Victim Model\\ Attack Method\end{tabular}} & \multicolumn{2}{c|}{BERT}                            & \multicolumn{2}{c|}{DistilBERT}                      & \multicolumn{2}{c|}{RoBERTa}    & \multicolumn{2}{c|}{BERT}                            & \multicolumn{2}{c|}{DistilBERT}                      & \multicolumn{2}{c|}{RoBERTa}    & \multicolumn{2}{c|}{BERT}                            & \multicolumn{2}{c|}{DistilBERT}                      & \multicolumn{2}{c}{RoBERTa}     \\ \cline{3-20} 
\multicolumn{1}{c|}{}                                  &                                                                                          & ASR            & \multicolumn{1}{c|}{CACC}           & ASR            & \multicolumn{1}{c|}{CACC}           & ASR            & CACC           & ASR            & \multicolumn{1}{c|}{CACC}           & ASR            & \multicolumn{1}{c|}{CACC}           & ASR            & CACC           & ASR            & \multicolumn{1}{c|}{CACC}           & ASR            & \multicolumn{1}{c|}{CACC}           & ASR            & CACC           \\ \midrule
\multicolumn{1}{c|}{\multirow{6}{*}{\makecell[c]{Low \\ Poison \\ Rate}}}  & Syntactic                                                                                & 51.59          & \multicolumn{1}{c|}{91.16}          & 54.77          & \multicolumn{1}{c|}{89.62}          & 46.71          & \textbf{93.52} & 50.17          & \multicolumn{1}{c|}{\textbf{92.00}} & 57.60          & \multicolumn{1}{c|}{\textbf{92.10}} & 70.67          & \textbf{91.40} & 80.96          & \multicolumn{1}{c|}{91.71}          & 84.87          & \multicolumn{1}{c|}{90.72}          & 87.77          & 91.21          \\
\multicolumn{1}{c|}{}                                  & Syntactic$_{aug}$                                                                              & 60.48          & \multicolumn{1}{c|}{\textbf{91.27}} & 57.41          & \multicolumn{1}{c|}{\textbf{90.39}} & 49.78          & 93.47          & 54.08          & \multicolumn{1}{c|}{91.85}          & 59.44          & \multicolumn{1}{c|}{91.90}          & 73.35          & 91.35          & 81.15          & \multicolumn{1}{c|}{\textbf{91.76}} & 84.19          & \multicolumn{1}{c|}{90.79}          & 91.37          & 91.18          \\
\multicolumn{1}{c|}{}                                  & Syntactic$_{mt}$                                                                               & \textbf{89.90} & \multicolumn{1}{c|}{90.72}          & \textbf{89.68} & \multicolumn{1}{c|}{89.84}          & \textbf{92.21} & 92.20          & \textbf{95.87} & \multicolumn{1}{c|}{91.80}          & \textbf{95.53} & \multicolumn{1}{c|}{91.30}          & \textbf{95.08} & 91.05          & \textbf{99.47} & \multicolumn{1}{c|}{\textbf{91.76}} & \textbf{99.26} & \multicolumn{1}{c|}{\textbf{91.25}} & \textbf{99.60} & \textbf{91.68} \\ \cline{2-20} 
\multicolumn{1}{c|}{}                                  & StyleBkd                                                                                 & 54.97          & \multicolumn{1}{c|}{91.16}          & 44.70           & \multicolumn{1}{c|}{90.50}          & 56.95          & \textbf{93.36} & 48.27          & \multicolumn{1}{c|}{\textbf{91.60}} & 48.27          & \multicolumn{1}{c|}{91.60}          & 58.32          & 90.40          & 69.62          & \multicolumn{1}{c|}{91.54}          & 71.41          & \multicolumn{1}{c|}{91.05}          & 64.86          & 91.07          \\
\multicolumn{1}{c|}{}                                  & StyleBkd$_{aug}$                                                                               & 58.28          & \multicolumn{1}{c|}{\textbf{91.98}} & 49.34          & \multicolumn{1}{c|}{\textbf{90.55}} & 58.72          & 92.59          & 49.66          & \multicolumn{1}{c|}{91.40}          & 49.16          & \multicolumn{1}{c|}{\textbf{92.10}} & 61.84          & \textbf{90.80} & 69.66          & \multicolumn{1}{c|}{\textbf{92.07}} & 73.21          & \multicolumn{1}{c|}{91.17}          & 63.81          & \textbf{91.50} \\
\multicolumn{1}{c|}{}                                  & StyleBkd$_{mt}$                                                                                & \textbf{83.44} & \multicolumn{1}{c|}{90.88}          & \textbf{81.35} & \multicolumn{1}{c|}{89.35}          & \textbf{89.07} & 92.81          & \textbf{78.88} & \multicolumn{1}{c|}{91.45}          & \textbf{74.41} & \multicolumn{1}{c|}{91.95}          & \textbf{84.25} & 90.60          & \textbf{92.40} & \multicolumn{1}{c|}{91.43}          & \textbf{93.95} & \multicolumn{1}{c|}{\textbf{91.18}} & \textbf{92.67} & 91.09          \\ \midrule
\multicolumn{1}{l|}{\multirow{4}{*}{\makecell[c]{Label\\Consistent}}} & Syntactic                                                                                & 84.41          & \multicolumn{1}{c|}{\textbf{91.38}} & 77.83          & \multicolumn{1}{c|}{\textbf{89.24}} & 70.61          & \textbf{92.59} & 93.02          & \multicolumn{1}{c|}{\textbf{88.95}} & 95.25          & \multicolumn{1}{c|}{\textbf{88.85}} & 98.49          & \textbf{89.35} & 70.14          & \multicolumn{1}{c|}{91.05}          & 62.67          & \multicolumn{1}{c|}{\textbf{90.66}} & 91.84          & 89.99          \\
\multicolumn{1}{l|}{}                                  & Syntactic$_{mt}$                                                                               & \textbf{94.40}  & \multicolumn{1}{c|}{90.72}          & \textbf{94.95} & \multicolumn{1}{c|}{89.13}          & \textbf{92.11} & \textbf{92.59} & \textbf{98.99} & \multicolumn{1}{c|}{88.74}          & \textbf{98.88} & \multicolumn{1}{c|}{88.69}          & \textbf{98.99} & 88.94          & \textbf{93.16} & \multicolumn{1}{c|}{\textbf{91.49}} & \textbf{99.46} & \multicolumn{1}{c|}{90.64}          & \textbf{99.28} & \textbf{90.42} \\ \cline{2-20} 
\multicolumn{1}{l|}{}                                  & StyleBkd                                                                                 & 66.00          & \multicolumn{1}{c|}{\textbf{90.83}} & 66.45          & \multicolumn{1}{c|}{\textbf{89.29}} & 73.07          & 92.53          & 61.96          & \multicolumn{1}{c|}{90.60}          & 59.39          & \multicolumn{1}{c|}{\textbf{90.60}} & 87.43          & \textbf{91.25} & 36.86          & \multicolumn{1}{c|}{\textbf{91.59}} & 35.81          & \multicolumn{1}{c|}{90.76}          & 42.08          & \textbf{90.76} \\
\multicolumn{1}{l|}{}                                  & StyleBkd$_{mt}$                                                                                & \textbf{84.99} & \multicolumn{1}{c|}{90.77}          & \textbf{85.21} & \multicolumn{1}{c|}{88.69}          & \textbf{91.50} & \textbf{92.81} & \textbf{83.63} & \multicolumn{1}{c|}{\textbf{91.10}} & \textbf{82.51} & \multicolumn{1}{c|}{90.40}          & \textbf{87.54} & 90.95          & \textbf{88.65} & \multicolumn{1}{c|}{91.58}          & \textbf{89.62} & \multicolumn{1}{c|}{\textbf{91.32}} & \textbf{92.78} & 90.14          \\ \bottomrule
\end{tabular}
}
\caption{\label{overall_results}	
Backdoor attack results in the low-poisoning-rate and label-consistent attack settings.
}

\end{table*}

%% file: tabs/clean-data.tex
\begin{table*}[t]
\centering
\resizebox{\textwidth}{!}{\begin{tabular}{c|c|cc|cc|cc|cc|cc|cc}
\toprule
\multirow{2}{*}{Dataset}     & \multirow{2}{*}{\begin{tabular}[c]{@{}c@{}}Victim Model\\ Attack Method\end{tabular}} & \multicolumn{2}{c|}{BERT}       & \multicolumn{2}{c|}{BERT-CFT}   & \multicolumn{2}{c|}{DistilBERT} & \multicolumn{2}{c|}{DistilBERT-CFT} & \multicolumn{2}{c|}{RoBERTa}    & \multicolumn{2}{c}{RoBERTa-CFT} \\ \cline{3-14} 
                             &                                                                                       & ASR            & CACC           & ASR            & CACC           & ASR            & CACC           & ASR              & CACC             & ASR            & CACC           & ASR            & CACC           \\ \midrule
\multirow{6}{*}{SST-2}       & Syntactic                                                                             & 97.91          & 89.84          & 70.91          & 92.09          & 97.91          & 86.71          & 67.40            & \textbf{90.88}   & 97.37          & 90.94          & 56.58          & \textbf{93.30} \\
                             & Syntactic$_{aug}$                                                                           & \textbf{99.45} & \textbf{90.61} & \textbf{98.90} & 90.10          & \textbf{99.67} & \textbf{88.91} & \textbf{96.49}   & 89.79            & 97.15          & \textbf{91.76} & \textbf{83.99} & 93.25          \\
                             & Syntactic$_{mt}$                                                                            & 99.12          & 88.74          & 85.95          & \textbf{92.53} & 99.01          & 85.94          & 78.92            & 90.00            & \textbf{98.25} & 91.38          & 74.12          & 93.03          \\ \cline{2-14} 
                             & StyleBkd                                                                              & 92.60          & 89.02          & 77.48          & \textbf{91.71} & 91.61          & \textbf{88.30} & 76.82            & 90.23            & 93.49          & 91.60          & 84.11          & \textbf{93.36} \\
                             & StyleBkd$_{aug}$                                                                            & 95.47          & \textbf{89.46} & \textbf{91.94} & 91.16          & \textbf{95.36} & 87.64          & \textbf{92.27}   & 88.91            & 94.92          & \textbf{91.98} & 85.32          & 92.97          \\
                             & StyleBkd$_{mt}$                                                                             & \textbf{95.75} & 89.07          & 82.78          & 91.49          & 94.04          & 87.97          & 84.66            & \textbf{90.50}   & \textbf{96.80} & 90.72          & \textbf{88.96} & 93.19          \\ \midrule
\multirow{6}{*}{Hate-Speech} & Syntactic                                                                             & 97.49          & 90.25          & 78.60          & 90.70          & 97.93          & 89.70          & 65.42            & \textbf{91.40}   & 99.27          & 90.45          & 85.47          & 91.70          \\
                             & Syntactic$_{aug}$                                                                           & 98.04          & \textbf{91.05} & \textbf{93.13} & 91.20          & 97.43          & \textbf{90.80}  & 86.98            & 91.05            & \textbf{99.32} & \textbf{91.35} & \textbf{98.21} & 91.60          \\
                             & Syntactic$_{mt}$                                                                            & \textbf{99.22} & 90.05          & 79.66          & \textbf{91.55} & \textbf{99.16} & 89.84          & \textbf{88.49}   & 91.15            & 98.83          & 89.84          & 94.92          & \textbf{91.80} \\ \cline{2-14} 
                             & StyleBkd                                                                              & 86.15          & 89.35          & 64.25          & \textbf{92.10} & 85.87          & 89.00          & 64.64            & 91.60            & 94.86          & 90.30          & 81.06          & 90.50          \\
                             & StyleBkd$_{aug}$                                                                            & 87.49          & \textbf{90.00} & 78.49          & 91.10          & 86.76          & \textbf{89.45} & \textbf{77.21}   & 91.10            & 99.22          & \textbf{91.10} & \textbf{95.53} & 90.95          \\
                             & StyleBkd$_{mt}$                                                                             & \textbf{91.01} & 89.14          & \textbf{78.72} & 91.60          & \textbf{90.78} & 87.79          & 71.34            & \textbf{91.70}   & \textbf{99.50} & 88.99          & 91.17          & \textbf{91.20} \\ \midrule
\multirow{6}{*}{AG's News}   & Syntactic                                                                             & 98.86          & \textbf{91.45} & 91.14          & \textbf{92.05} & 99.26          & 90.68          & 89.59            & \textbf{91.28}   & \textbf{99.53} & 90.45          & 96.30          & \textbf{91.43} \\
                             & Syntactic$_{aug}$                                                                           & 99.07          & \textbf{91.45} & 91.44          & 91.72          & 99.28          & \textbf{91.04} & 93.31            & 91.13            & 99.47          & \textbf{91.22} & 98.28          & 91.34          \\
                             & Syntactic$_{mt}$                                                                            & \textbf{99.79} & 91.28          & \textbf{97.16} & 91.74          & \textbf{99.82} & 90.75          & \textbf{97.77}   & 90.84            & 99.47          & 90.43          & \textbf{98.96} & 91.03          \\ \cline{2-14} 
                             & StyleBkd                                                                              & 96.59          & 90.39          & 82.35          & \textbf{91.88} & 96.49          & 89.67          & 80.84            & 91.26            & 96.28          & 89.68          & 78.92          & \textbf{91.37} \\
                             & StyleBkd$_{aug}$                                                                            & 96.25          & \textbf{91.05} & \textbf{86.91} & 91.64          & 96.73          & \textbf{89.80} & 81.79            & 91.17            & 96.19          & \textbf{89.99} & \textbf{91.81} & 90.78          \\
                             & StyleBkd$_{mt}$                                                                             & \textbf{98.00} & 90.17          & 84.77          & 91.64          & \textbf{97.64} & 89.49          & \textbf{90.69}   & \textbf{91.39}   & \textbf{98.18} & 89.22          & 82.91          & 91.21          \\ \bottomrule
\end{tabular}
}

\caption{\label{clean-data-finetune}	
Backdoor attack results in the setting of clean data fine-tuning.
}

\end{table*}

%% file: tabs/sub_exp.tex
\begin{table}[]
\centering
\resizebox{.49\linewidth}{!}{
\begin{tabular}{c|r}
\toprule
 Attack Method              & \multicolumn{1}{c}{Acc} \\ 
 \midrule
 Syntactic                     & 89.02                   \\
                       Syntactic$_{aug}$        & 92.54                   \\
                       Syntactic$_{mt}$            & \textbf{98.02}                   \\ 
\cline{1-2} 
                       StyleBkd            & 85.07                   \\
                       StyleBkd$_{aug}$ & 86.89                   \\
                       StyleBkdc$_{mt}$       & \textbf{94.14}                   \\ 
\bottomrule
\end{tabular}
}
\caption{\label{sub_exp} Probing accuracy on SST-2 of BERT.}

\end{table}

%% file: conclusion.tex
\section{Conclusion}
We present two simple tricks based on multi-task learning and data augmentation, respectively to make current backdoor attacks more harmful.
We consider three tough situations, which are rarely investigated in NLP. Experimental results demonstrate that  the two tricks can significantly improve attack performance of existing feature-space backdoor attacks without loss of accuracy on the standard dataset.
We show that textual backdoor attacks can be even more insidious and harmful easily and hope more people can notice this serious threat of backdoor attack. 
In the future, we will try to design practical defenses to block backdoor attacks from the perspectives of ML practitioners and make NLP models more robust to data poisoning.
